\input epsf
\documentclass[nohyper,notoc]{article} 
\usepackage{axodraw}
\usepackage{epsfig}

\def\beq{\begin{equation}}
\def\eeq{\end{equation}}
\def\bea{\begin{eqnarray}}
\def\eea{\end{eqnarray}}
\def\bq{\begin{quote}}
\def\eq{\end{quote}}
\def \lsim{\mathrel{\vcenter
     {\hbox{$<$}\nointerlineskip\hbox{$\sim$}}}}
\def \gsim{\mathrel{\vcenter
     {\hbox{$>$}\nointerlineskip\hbox{$\sim$}}}}
\def\gappeq{\mathrel{\rlap {\raise.5ex\hbox{$>$}}
{\lower.5ex\hbox{$\sim$}}}}
\def\lappeq{\mathrel{\rlap{\raise.5ex\hbox{$<$}}
{\lower.5ex\hbox{$\sim$}}}}

\def\Huv{\langle H_u \rangle}

\def\mnu{[m_{\nu}]_{\alpha \beta}}

\def\bea{\begin{eqnarray}}   
\def\eea{\end{eqnarray}}

\begin{document}
\vspace*{-1in}
\renewcommand{\thefootnote}{\fnsymbol{footnote}}
\begin{flushright}
LYCEN 2005-18\\
 IPPP/05/23\\
DCPT/05/46\\
FTUV/05-0608, IFIC/05-0608\\
%\texttt{hep-ph/yymmddd}
\end{flushright}
\vskip 5pt
\begin{center}
{\Large {\bf From transition magnetic moments to majorana
neutrino masses
}}
\vskip 25pt
{ Sacha Davidson $^{1,2}$\footnote{E-mail address: 
s.davidson@ipnl.in2p3.fr},
  Martin Gorbahn  $^{2}$ \footnote{E-mail address: 
martin.gorbahn@durham.ac.uk },
 and Arcadi Santamaria $^{3}$ \footnote{E-mail 
address:Arcadi.Santamaria@uv.es }} 
\vskip 10pt 
$^1${\it IPNL, Universit\'e C.B. Lyon-1, 4,rue Enrico Fermi, 
Villeurbanne, cedex 69622 France} \\ %$~$\\
$^2${\it IPPP
   University of Durham,
%South Road,
Durham DH1 3LE, U.K. }\\ %$~$\\
$^3${\it Dept.\ de F\'{\i}sica Te\`orica and IFIC, Univ. Valencia-CSIC, 
Dr. Moliner 50, 46100 Burjassot, Spain }

\vskip 20pt
{\bf Abstract}
\end{center}

\begin{quotation}
  {\noindent\small 
It is well known that a majorana mass induces a (small)
transition magnetic moment. The converse is also true;
in this paper we estimate the loop contribution
of transition magnetic moments $[\mu]_{\alpha \beta}$
 to the neutrino mass matrix $[m]_{\alpha \beta}$.
We  show that for hierarchical neutrino masses,
 %and
%magnetic moment flavour choices, 
the contribution of $[\mu]_{e \tau}$ to  $[m]_{e \tau }$
can exceed the experimental value of  $[m]_{e \tau }$.
\vskip 10pt
\noindent
%PACS number(s):~12.60.Jv, 14.60.Pq, 11.30.Fs
}

\end{quotation}
%\noindent{Neutrino Physics,
%Supersymmetric Standard Model, Solar and Atmospheric Neutrinos}

\vskip 20pt  

\setcounter{footnote}{0}
\renewcommand{\thefootnote}{\arabic{footnote}}

%%%%%%%%%%%%%%%%%%%%%%%%%%%%%%%%%%%%%%%%%%%%%%%%%%%%%%%%%%%%%%%%%%%%%%
%% INTRO        %%%%%%%%%%%%%%%%%%%%%%%%%%%%%%%%%%%%%%%%%%%%%%%%%%%%%%
%%%%%%%%%%%%%%%%%%%%%%%%%%%%%%%%%%%%%%%%%%%%%%%%%%%%%%%%%%%%%%%%%%%%%%
%\newpage
\section{Introduction}
\label{intro}

%\subsection{motivation}

If  neutrino masses are majorana, they are generated
by some New Physics(NP) from beyond the Standard Model (SM).
There is a plethora of models that  
fit the low energy  neutrino mass matrix,
%explain
%the neutrino masses and mixing angles 
%(which is measured at the electroweak
%scale or below), 
but that differ in interactions and particle
content at higher energies. 
This motivates the question
``can we identify this new physics from  data?''---
preferably from experiments in our lifetime, which
suggests that they should be performed at energy
scales within an order of magnitude or so of $m_W$. 

If the scale of the new physics is experimentally
accessible (eg R-parity violating supersymmetry), 
then the answer is ``yes''. But
if the scale of new physics is $above$ that
accessible to accelerators, we  are lead to ask
``to what degree can we
reconstruct the new physics of the lepton sector by
measuring the coefficients of non-renormalizable operators''?

This is an complex question. This paper 
considers a more manageable toy model, restricted
to lepton number violating operators:
the neutrino masses are majorana, and  we suppose that
neutrino transition magnetic moments \cite{ggr} 
are of order their current upper bound. What can be learned
from the effective theory\cite{effth} comprising these non-renormalizable
operators and the Standard Model? We find constraints on
transition magnetic moment operators, from their
contribution to majorana masses. The effective Lagrangian 
point of view will allow us to derive  these bounds 
in a model independent way.

In the remainder of this section, we discuss 
diagrams that could generate a majorana mass from a 
magnetic moment.
In section  \ref{notn}, we review our notation,  current
bounds on neutrino masses and magnetic moments, the sense
in which   magnetic  moments near
their current bound are ``large'' but
masses are ``small'', and
models built to address such an unexpected occurrence.
The transition magnetic moment operator is of dimension
7 in the electroweak sector of the SM. Above the scale
$m_W$, and below  $\Lambda_{NP}$ (the scale of the new physics
which generates the masses and/or magnetic moments),
there should be a range in energy where the operator
evolves according to  the renormalization group
of the SM. So in section \ref{RG}, we compute
the anomalous dimensions of  the relevant
lepton number violating operators, which 
gives the leading contribution of the magnetic moment
operator to the neutrino masses.
For some parameter choices, this contribution is large.
In section \ref{sowhat}, we  discuss the implications
of this calculation.

\subsection{diagrams} 

%%%%%%%%%%%%%%%%%%%%%%%%%%%%%%%%%%%%%%%%%%%%%%%%%%%%%%%%%%%%%%%%%%%%%%
%% SUBSECT DIAGRAMS  %%%%%%%%%%%%%%%%%%%%%%%%%%%%%%%%%%%%%%%%%%%%%%%%%%%%%%
%%%%%%%%%%%%%%%%%%%%%%%%%%%%%%%%%%%%%%%%%%%%%%%%%%%%%%%%%%%%%%%%%%%%%%
\label{diagrams}

If the light neutrinos are majorana, they can have transition
magnetic moments.
Like the neutrino majorana mass, the magnetic moment  $[\mu]_{\alpha \beta}$ 
violates
lepton number by two units. So a diagram connecting the photon line
back to either of the neutrino lines would appear to 
contribute to the majorana mass---except that
it is negligibly small ($O(\mu^2)$), because the only
$ \nu_\alpha \nu_\beta \gamma$ interaction in a low energy ($\ll m_W$)
U(1)$_{em}$-invariant Lagrangian
is $[\mu]_{\alpha \beta}$.  Furthermore, 
this $O(\mu^2)$  diagram would be $L$ conserving,
so cannot contribute to majorana masses.

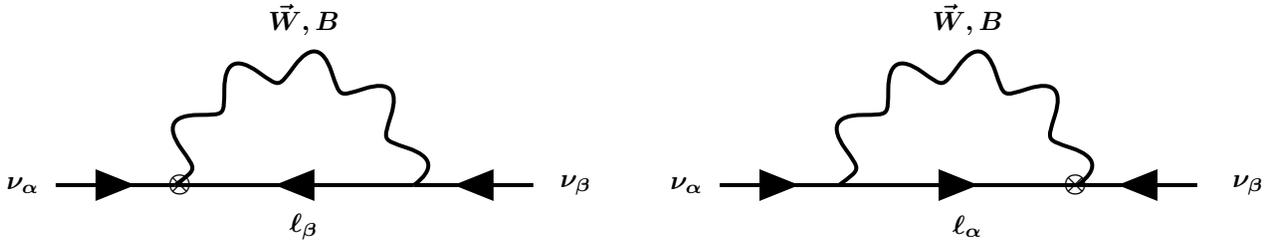
\begin{figure}[ht]
\unitlength1.1mm
\SetScale{3}
\begin{boldmath}
\begin{center}
\begin{picture}(60,30)(0,0)
\ArrowLine(0,0)(15,0)
\ArrowLine(45,0)(15,0)
\ArrowLine(60,0)(45,0)
\PhotonArc(30,0)(15,0,180){2}{5}
%\Photon(40.6,10.6)(60,30){2}{5}
\Text(-2,0)[r]{$\nu_\alpha$}
\Text(61,0)[l]{$\nu_\beta$}
\Text(15,0)[c]{$\otimes$}
\Text(30,20)[c]{$\vec{W},B$}
\Text(30,-5)[c]{$\ell_\beta$}
\end{picture}
\qquad\qquad\qquad
\begin{picture}(60,30)(0,0)
\ArrowLine(0,0)(15,0)
\ArrowLine(15,0)(45,0)
\ArrowLine(60,0)(45,0)
\PhotonArc(30,0)(15,0,180){2}{5}
%\Photon(38,20)(60,0){-2}{5}
\Text(-2,0)[r]{$\nu_\alpha$}
\Text(62,0)[l]{$\nu_\beta$}
\Text(43,0)[c]{$\otimes$}
\Text(30,20)[c]{$\vec{W},B$}
\Text(30,-5)[c]{$\ell_\alpha$}
\end{picture}
\end{center}
\end{boldmath}
\vspace{10mm}
\caption{One-loop diagrams mediated by SU(2) gauge bosons,
whose combined contribution to $[m_\nu]_{\alpha \beta}$ sums to zero.
The crossed vertex is the magnetic moment $[\mu]_{\alpha \beta}$,
and $\ell$ is the lepton doublet.}
\label{oneloop}
\end{figure}

However, the magnetic moment operator  
must be  the $E \ll \Huv$ realisation of an $SU(2) \times U(1)$ invariant
operator, so there  should be a related 
$\nu \nu Z$ interaction,  and/or possibly 
a  $ \nu e W^+$ interaction.
See section \ref{notn}
for a discussion of  $SU(2) \times U(1)$ invariant
operators which  reduce to a neutrino transition magnetic moment.
 Replacing the photon with a $Z$, and connecting the $Z$
back to either of the neutrino lines,  as in fig. \ref{oneloop},
gives a pair of one-loop diagrams contributing to $\mnu$---which
sum to zero. This is expected: $\mnu$ is symmetric on
generation indices, and $[\mu]_{\alpha \beta}$ is anti-symmetric, so 
$[\mu]_{\alpha \beta}$ must be multiplied by  some flavour-antisymmetric 
matrix to contribute to $\mnu$. This feature
was pointed out in \cite{Vol}, and used
in models that produce small majorana masses
and large magnetic moments\cite{others,BFZ}.

\begin{figure}[ht]
\unitlength1.1mm
\SetScale{3}
\begin{boldmath}
\begin{center}
\begin{picture}(60,30)(0,0)
\ArrowLine(0,0)(15,0)
\ArrowLine(45,0)(15,0)
\ArrowLine(60,0)(45,0)
\PhotonArc(30,0)(15,0,180){2}{5}
%\Photon(40.6,10.6)(60,30){2}{5}
\Text(-2,0)[r]{$\nu_\alpha$}
\Text(61,0)[l]{$\nu_\beta$}
\Text(15,0)[c]{$\otimes$}
\Text(30,20)[c]{${W}^{+}$}
\Text(30,-7)[c]{$e_{R \beta}$}
\Text(20,0)[c]{$X$}
\Text(38,0)[c]{$X$}
\Text(20,-5)[c]{$m^e_\beta$}
\Text(38,-5)[c]{$m^e_\beta$}
\end{picture}
\qquad\qquad\qquad
\begin{picture}(80,50)(-15,-20)
\ArrowLine(-15,0)(25,0)
\ArrowLine(25,0)(45,0)
\ArrowLine(60,0)(45,0)
\PhotonArc(15,0)(15,180,360){2}{5}
\PhotonArc(30,0)(15,0,180){2}{5}
%\Photon(40.6,10.6)(60,30){2}{5}
\Text(-17,0)[r]{$\nu_\alpha$}
\Text(61,0)[l]{$\nu_\beta$}
\Text(43,0)[c]{$\otimes$}
\Text(30,20)[c]{$B$}
\Text(15,-20)[c]{$W^+$}
\Text(37,-5)[c]{$\nu_\alpha$}
\Text(10,3)[c]{$e_{ \alpha}$}
\Text(20,0)[c]{$X$}
\Text(24,0)[c]{$X$}
\Text(22,-5)[c]{$(m^e_\alpha)^2$}
%\Text(38,-5)[c]{$m^e_\alpha$}
\end{picture}
\end{center}
\end{boldmath}
\vspace{10mm}
\caption{On the left, a one-loop contribution
to $[m_\nu]_{\alpha \beta}$, where
the crossed vertex is a magnetic moment $[\mu]_{\alpha \beta}$
involving the SU(2) gauge bosons $\vec{W}$, and $X$ is 
a mass insertion. 
On the right, a
two-loop  diagram, 
the lowest order  contribution  when the
magnetic moment $[\mu]_{\alpha \beta}$
involves the hypercharge gauge boson $B$.
}
\label{nonzero}
\end{figure}
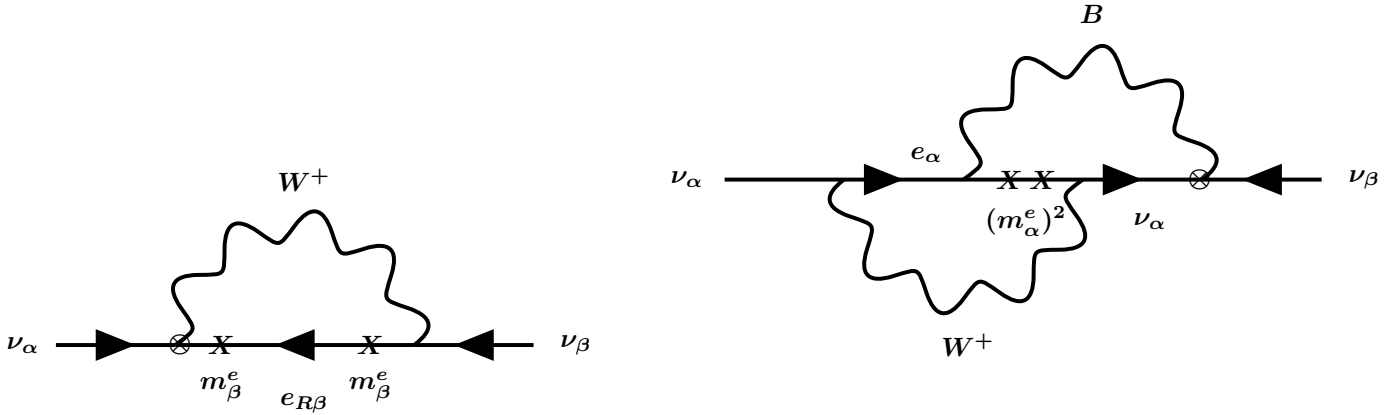

The required antisymmetric matrix could be
$(m^e_\beta)^2 -(m^e_\alpha)^2 $, where $m^e_\alpha$ is a charged lepton
mass.  In the one loop diagram
that includes  the neutral $Z$, the remaining loop particles must also be
neutral, so $m^e_\alpha$ cannot appear.
However, if the magnetic moment operator induces a $\nu e W$
interaction,then  the diagram on the left of fig. \ref{nonzero}
can give a one-loop  contribution to $\mnu$ 
 of order
\beq
\frac{g \mu_{\alpha \beta} |m^{e2}_\alpha - m^{e2}_\beta|}
{16 \pi^2} \log \frac{\Lambda_{NP}^2}{m_W^2} 
%\gappeq  10^{-2} \frac{ a_{\alpha \beta}}{10^{-12} \mu_B}
%\frac{ |m^{e2}_\alpha - m^{e2}_\beta|}{m_\tau^2}   ~~eV  ~~~~~~(1-loop)
\label{estimate-1}
\eeq
which is $\gsim 10^{-2}$ eV for $[\mu]_{\tau \alpha} \simeq 10^{-12} \mu_B$.
Notice that by power counting, 
this diagram is logarithmically divergent, {\it i.e.} there
are no
quadratic divergences. This is because the magnetic 
moment is  antisymmetric
in flavour, while neutrino masses are symmetric, so
there are two charged lepton mass insertions.

Alternatively, if $[\mu]$ only involves the $\gamma$ and $Z$,
then two loops are required to have charged particles
in the diagram.  A two-loop contribution to $\mnu$  is
illustrated on the right in fig. \ref{nonzero}.
The  $m^e_\alpha$ must appear twice on the
charged lepton line, because the $W$ only
interacts with left-handed fermions. This diagram is log divergent, and 
should be of order
\beq
\frac{g^3 \mu_{\alpha \beta} |m^{e2}_\alpha -m^{e2}_\beta| }{
(16 \pi^2)^2} \log^2 \frac{\Lambda_{NP}^2}{m_W^2} 
~~~~~~~~~({\rm 2-loop}).
\label{estimate0}
\eeq

The one-loop $W$ diagram 
will give the largest contribution to the mass
matrix---if the $W e \nu$ interaction exists.
This   is theoretically ``reasonable'': 
inside the crossed effective vertex of figures
\ref{oneloop} and \ref{nonzero} are loops that
generate $[\mu]_{\alpha \beta}$.  Some of the particles 
in  these loops 
must carry SU(2) quantum numbers, because there is no
renormalizable vertex involving $\nu \nu$ and an SU(2) singlet.
So if the effective vertex generating $[\mu]_{\alpha \beta}$ is opened
up and one looks at the constituent diagrams, 
naively it seems  that a $W$ magnetic  moment operator could
arise by attaching  $W$ to an internal line. 
% one-loop calculations are easier than two.
In section \ref{RG}, we will calculate the 
 $W$-magnetic moment contribution
to the  neutrino mass matrix, and estimate
the two-loop   $Z$ contribution.

\section{notation, bounds and expectations}
\label{notn}

Suppose the light neutrinos are majorana. Then it is known \cite{Nieves:1981zt}
that they can have transition magnetic moments, but not flavour diagonal
ones.
This is because the magnetic  moment interaction, which we
normalize in the Lagrangian as:
\bea
\frac{\mu_{\alpha \beta}}{2} \overline{\psi}_\alpha \sigma^{\mu \nu} \psi_\beta F_{\mu \nu} 
& \rightarrow &
\frac{\mu_{\alpha \beta}}{2} 
 ~ \overline{\nu^c}_{\alpha} \sigma^{\mu \nu}
P_L \nu_{\beta} ( F_{\mu \nu}) + h.c.
\label{op}
\eea
flips the chirality  of the fermion passing through. 
Greek indices from the beginning of the alphabet, $e.g.$ 
$\alpha, \beta$,  are flavour indices,  $\psi$ is a four-component
fermion,
$\overline{\nu^c} = (-i \gamma_2 (\nu ^\dagger)^T)^\dagger \gamma_0$,
 and the magnetic moment  $[\mu]_{\alpha \beta}$ has dimensions
of 1/mass.  Matrices in flavour space will
often be written in square brackets, {\it e.g.} $[\mu]$ and $[m_\nu]$. 
 The  ``right-handed'' component of  a light majorana 
neutrino is the antineutrino, so in chiral four-component notation for $\nu$, 
the left hand side  of eqn (\ref{op}) can be
rewritten as the right hand side, which vanishes for $ \alpha = \beta$
( $[\mu]_{\alpha \beta} = - [\mu]_{\beta \alpha}$) by
antisymmetry of fermion interchange.

The operator (\ref{op}) is of mass dimension five,
and consistent with electromagnetic gauge invariance. 
However it carries hypercharge $Y = -2$,
so the   SU(2)$ \times$ U(1) invariant
operator must be of higher dimension involving two Higgses.
There are two possible dimension seven operators
which give a neutrino magnetic moment interaction after 
spontaneous symmetry breaking:
\beq
[O_B]_{\alpha \beta}
= {g'} (\overline{\ell^c}_\alpha \epsilon H)\sigma^{\mu \nu} 
(H \epsilon P_L \ell_\beta) B_{\mu \nu} ,
~~~~ [O_W]_{\alpha \beta} = i {g} \varepsilon_{abd} 
(\overline{\ell^c}_\alpha \epsilon  \tau^a P_L \ell_\beta)
( H \epsilon  \tau^b H) W_{\mu \nu}^d .
\label{magmoSM}
\eeq
where
%$\overline{\ell^c} = (-i \gamma_2 (\ell ^\dagger)^T)^\dagger \gamma_0$, 
the lepton flavour \footnote{we can work
in the flavour basis for the neutrinos, because the
neutrino masses are small compared to any relevant energy
scale, so will induce a negligible correction. The MNS matrix
therefore will not appear in our calculation}
indices $\alpha, \beta  \in \{e,\mu,\tau \}$ are explicit, $\{\tau_i \}$
are the  SU(2) Pauli matrices, the SU(2) contractions are
implicit in the parentheses
($\epsilon  = - i \tau_2$, $(v \epsilon u) = v_2 u_1 - v_1 u_2)$,
$\varepsilon_{abd} \neq \epsilon$ is the totally antisymmetric tensor,
and $W_{\mu \nu}, B_{\mu \nu}$ are the gauge field strength tensors
for SU(2) and U(1)$_Y$. We define the operators 
without hermitian conjugates;  in the Lagrangian,
they will appear  multiplied by Wilson coefficients and $+ h.c.$
---see, {\it e.g.} eqn   (\ref{eq:effective_hamiltonian}).

We will compute the magnetic moment contribution to the
masses by renormalization group mixing, for which we 
need the dimension seven mass term 
that arises with a single Higgs doublet \cite{babu}:
\beq
[O_M]_{\alpha \beta}  = 
(\overline{\ell^c}_\alpha \epsilon H) (H \epsilon P_L \ell_\beta) 
(H ^\dagger H) ~~~.
\label{OM}
\eeq
In the presence of more than one
Higgs doublet, there are other interesting operators which
lead to neutrino masses \cite{aj,cei}. 

\subsection{Phenomenological bounds}

A $\Delta L = 2$  coupling between a gauge boson and a pair
of leptons, of different flavour, could have various
observable effects. We did not find significant
bounds from rare decays ({\it e.g.}  $W \nu e$ couplings),
or precision lepton number conserving processes
like $g-2$.   The 
$W e \nu_{\mu , \tau} $ interaction could appear at one of the vertices
in neutrinoless double  beta decay, but  
is not significantly constrained because of
the flavour antisymmetry.  The strongest bounds are
on the magnetic moment interaction between a photon and a pair
of neutrinos. 
This allows radiative decays $\nu_j \rightarrow
\bar{\nu}_i \gamma$,   contributes to the $\nu -e$ scattering
cross-section, and induces the ``decay'' of photons
in a plasma into $\nu$ pairs. Lower bounds on
the neutrino lifetime do not set interesting constraints
on the magnetic moments, because the decay rates are already
suppressed by powers of $m_\nu$. 
Bounds on $[\mu]_{ \ell \beta}$  from $\nu$ scattering
experiments,  are \cite{PDB} 
\beq
2 \mu_{e\beta} \leq 0.9 \times 10^{-10} \mu_B,~~~
2 \mu_{\mu \beta} \leq 6.8 \times 10^{-10}\mu_B,~~~
2 \mu_{\tau \beta} \leq 3.9 \times 10^{-7} \mu_B ~~~~~~~{\rm expt}
\label{bdexpt}
\eeq
where $\mu_B = e/(2 m_e)$, and the 2 is because
our neutrinos are majorana \cite{ggr}. For transition magnetic
moments, $[\mu]_{\tau \beta} $ must satisfy the
bound on $[\mu]_{e \tau} $ or $[\mu]_{ \mu \tau} $, 
so is $\leq 6.8 \times 10^{-10} \mu_B$.
The most restrictive constraint on  $[\mu]_{\alpha \beta}$  
comes from astrophysics. If photons in a
stellar plasma  can ``decay'' $\gamma \rightarrow \nu_\alpha \nu_\beta$,
their energy escapes the star immediately.  The observed
cooling rate of globular cluster stars therefore
sets a bound \cite{ggr99}
\beq
2 [\mu]_{\alpha \beta} \lappeq 3 \times 10^{-12} \mu_B~~~~~~~{\rm astro~.}
\label{bdastro}
\eeq

The solar neutrino flux is explained by  large mixing angle (LMA)
oscillations,
but a subdominant effect due to neutrino magnetic moments
remains possible. In the presence of the solar
magnetic field,  a non-zero $[\mu]_{e \alpha}$ 
could precess\footnote{see, $e.g.$ \cite{ggr}
for an introduction to spin-flavour precession, and
\cite{af} for up-to-date references.}
 the  $\nu_e$  into $\bar{\nu}_\alpha$s .
However, the solar bound
on $[\mu]_{e \alpha}$ \cite{mariana} and  the sensitivity 
of future experiments  are somewhat unclear, both
theoretically and experimentally. 
Spin-flavour precession is usually assumed to
take place in the outer regions of the sun, but a recent
analysis \cite{af} suggests that this is not
the case for $^8 Be$ neutrinos with LMA parameters.
In addition, the effect depends  on  
the solar magnetic field. In the accumulated solar
neutrino data,   there is 
some evidence for time-dependence. 
However, the longterm  
anti-correlation found between the Homestake
solar $\nu$ data  and the solar cycle
\cite{Davis} is not found by  the SK collaboration \cite{SKnot}
(but see \cite{but}), which
also does not find evidence for anti-neutrinos \cite{SK2}. 
The sensitivity of 
 future solar neutrino data
to neutrino magnetic moments
was studied in   \cite{CV} (see also \cite{af});  we cavalierly
extract that  $[\mu]_{e \alpha}$  in the range
$10^{-10} \rightarrow 10^{-13} \mu_B$ could
be interesting, and for
all our numerical estimates, we will take 
\beq
[\mu]_{\alpha \beta} \simeq % \times 
10^{-12} \mu_B ~~.
\label{number}
\eeq

\subsection{Dimensional analysis}

Dimensional
analysis suggests that $m_\nu \sim .1 eV$ is ``small'', whereas
$\mu \sim 10^{-12} \mu_B$ is ``large''. That is, 
new lepton number non-conserving physics
at some scale $M$  could induce
 transition magnetic moments and/or majorana masses;
$M$  estimated 
from  $m_\nu$ is significantly higher than that obtained from $\mu$.

It is well known, that if the  dimension five  majorana mass operator 
$(H \ell)(H \ell)$ 
induces  neutrino masses $ m_\nu \sim .1 eV$
then the New Physics scale where this operator is generated
should be $ \lsim v^2/(.1 eV) \sim 10^{14}$ GeV. 
To repeat this argument for the dimension seven
 transition magnetic  moment   operators  of
section \ref{notn},  %, with coefficients $C_i$. 
requires estimating a lower bound on their
coefficients $C_J$: 
\beq
\mu \sim  C_J v^2 \sim \frac{ {\cal B} }{M^3} v^2
\sim 10^{-12} \mu_B
\label{adima}
\eeq
where  ${\cal B} $ is  some combination of coupling constants 
and $1/(16 \pi^2)$ for loops,  and $M$ is
the mass scale of the diagram. 
The  magnetic moment must be suppressed by a loop factor
${\cal B}  \lsim g^2/(16 \pi^2)$,  because all
its external legs are neutral, but nonetheless the
photon should couple. 
Setting  $  {\cal B}  v^2$ as large
as is ``reasonable''  $ \sim  m_W^2/(8 \pi^2)$, gives 
\beq
M^3 \lsim 5 \times 10^{11} GeV^3 \left( \frac{10^{-12} \mu_B}{\mu} \right)
\label{almost}
\eeq
or $M \lsim 10$ TeV, if it is the same mass scale cubed.
Taking $M^3 \sim m_W^2 M_{max}$, to maximise the New Physics scale,
gives $M_{max} \lsim 10^8$ GeV.
If new ($\Delta L = 2$) physics arises between $10^4$ and
$10^7$ GeV, then the  $(H \epsilon \ell) (H \epsilon \ell)$
operator could  give  $m_\nu $  as large as   GeV to MeV. 
The coefficient of this dimension five 
operator, which is determined by the  New Physics, must therefore
be strongly suppressed.

Notice that the magnetic moment is measured in the  units used for
the electron magnetic moment, $\mu_B = e/(2m_e)$. This
is the relevant dimension for the electron, because
the momentum in its loops (contributing, for instance,
to $g-2$) is $1/p^2 \sim 1/m_e^2$, and $m_e$ must
appear upstairs to flip chirality. However, for
the weakly interacting neutrino, one might expect
$1/p^2 \sim 1/m_W^2$, suppressing 
 $ \mu \sim (m_e^2/m_W^2) \mu_B$.
So  the numerical value of eqn (\ref{number})  suggests
lepton number violation near the weak scale.

Various
models have been constructed,  which ``naturally'' generate a large
$[\mu]$ with small $\mnu$.
From a top-down perspective, the difficulty is %that
% If one looks 
``inside'' the 
magnetic moment vertex of figures \ref{oneloop}
and \ref{nonzero}, where  the new physics 
generates $\mu$.   If
 the photon is removed from these internal diagrams, 
it would naively seem that   
the dimension five neutrino mass operator is obtained,
with a ``natural'' coefficient of order the inverse new physics scale.
This  is too large.
%diagrams  contributing
%a large  value to $m_\nu$ are obtained.
Voloshin \cite{Vol}  %constructed
addressed this in  a 
model, %with large magnetic moments
%and small masses, 
by observing that $[\mu]_{\alpha \beta}$ was flavour {\it antisymmetric},
and  arranging cancellations
among the diagrams contributing to the 
flavour {\it symmetric} mass matrix.
This approach has been followed by many people \cite{others},
who exploit the flavour antisymmetry of $\mu$, and impose
additional symmetries on the New Physics,
to suppress  contributions to the neutrino masses. 
Another interesting model \cite{BFZ},
%is  such that
forbids  by angular momentum conservation
the magnetic moment diagram with its photon removed.

%%%%%%%%%%%%%%%%%%%%%%%%%%%%%%%%%%%%%%%%%%%%%%%%%%%%%%%%%%%%%%%%%%%%%%
%% SECT  OPERATOR RUNNING  %%%%%%%%%%%%%%%%%%%%%%%%%%%%%%%%%%%%%%%%%%%
%%%%%%%%%%%%%%%%%%%%%%%%%%%%%%%%%%%%%%%%%%%%%%%%%%%%%%%%%%%%%%%%%%%%%%

\section{leading logarithmic contributions}
\label{RG}

In this section, we estimate the leading logarithmic contribution of
the magnetic moment operators to $\mnu$. We first study the
 scenario  where the $O_W$ operator gives the
dominant contribution to the mixing into the 
dimension seven neutrino mass operator,
while in the second scenario the  New Physics  induces  the  $O_B$
operator, but not $O_W$.
We work at one loop in  unbroken  $SU(2) \times U_Y(1)$,
so  the propagating particles in our diagrams are massless,
and  the charged fermion mass insertions are replaced by
Yukawa couplings.   Then, the dimension
seven magnetic moment operator  cannot mix to the dimension 5
neutrino mass operator.

Let us suppose that new  (lepton number non-conserving) physics,
above the scale $\Lambda_{NP}$, can be matched onto   an effective theory
\begin{equation}
  \label{eq:effective_hamiltonian}
  H_{\rm eff}(\Lambda_{\rm NP}) =
  C^W_{\alpha \beta}(\Lambda_{\rm NP}) O^W_{\alpha \beta} +
  C^B_{\alpha \beta}(\Lambda_{\rm NP}) O^B_{\alpha \beta} +
  C^M_{\alpha \beta}(\Lambda_{\rm NP}) O^M_{\alpha \beta} +
  \ldots + h.c. = \vec{Q}^T \vec{C} (\Lambda_{\rm NP}) + h.c.
\end{equation}
where the operators are defined in eqns (\ref{magmoSM}) and
(\ref{OM}).
 Below the scale $\Lambda_{NP}$ our theory contains the
interactions and particles of the electroweak standard model,  the
lepton number violating non-renormalizable operators of Eq.
(\ref{eq:effective_hamiltonian}), and the
dimension five neutrino mass operator,  with
small coefficient as discussed
after eqn (\ref{almost}). This allows us to calculate the
contribution of the neutrino magnetic moment to the neutrino mass in a
way independent of the new physics scenario considered. To do this we
solve the renormalization group equation
\begin{equation}
\label{eq:operator_rge}
\mu \frac{d}{d \mu} \vec{C} (\mu) = 
\hat{\gamma}^T (g) \vec{C}  (\mu) % (\Lambda_{\rm NP})
\end{equation}
($\mu$ without indices or brackets being the renormalization scale)
by expanding the anomalous dimension
\begin{equation}
  \hat{\gamma} (g) = 
  \sum^\infty_{i = 0} \left ( \frac{g^2}{16 \pi^2} \right )^{i + 1} \hat{\gamma}^{(i)}
\end{equation}
and the Wilson coefficients 
\begin{equation}
  \vec{C} (\mu) = \sum^\infty_{i = 0} \left (
\frac{g^2}{16 \pi^2} \right )^{i} \vec{C}^{(i)}(\mu) 
\end{equation}
in terms of the weak coupling constant.
If we expand up to the second logarithmic enhanced order,
the terms  that will be relevant for us  are
%If we expand up to the second logarithmic enhanced order,
%we can solve eqn (\ref{eq:operator_rge}) perturbatively 
%
\begin{equation}
  \vec{C}(\mu) = 
  \vec{C}^{(0)}(\Lambda_{\rm NP}) +
    \frac{1}{2} \hat{\gamma}^{(0)^T} \vec{C}^{(0)}(\Lambda_{\rm NP})
    \frac{g^2}{16 \pi^2} \log \frac{\mu^2}{\Lambda_{\rm NP}^2} +
    \frac{1}{8} \hat{\gamma}^{(0)^T} \hat{\gamma}^{(0)^T} 
    \vec{C}^{(0)}(\Lambda_{\rm NP})
    \left(
      \frac{g^2}{16 \pi^2} \log \frac{\mu^2}{\Lambda_{\rm NP}^2}
    \right)^2.
\label{martin1}
\end{equation}

Suppose first that the New Physics generates
the operator  $O_W$ (so we neglect  $O_B$).
Then only the first term from eqn (\ref{martin1}) is required,
with $g^2(\mu) = g^2$,  and the $\hat{\gamma}^{(0)}$ matrix element
mixing $O_W$ to $O_M$. 
The self mixing of the magnetic moment operators can be neglected,  
so the Wilson coefficient (at $\Lambda_{NP}$)  is matched  onto the 
neutrino magnetic moment
\begin{equation}
 C^W_{\alpha \beta}(\Lambda_{\rm NP}) = - 
\frac{[\mu]_{\alpha \beta}}{4  v^2  g \sin \theta_W}
\end{equation}
where $v = \langle H \rangle$. 
Using the $W^+ e \nu$ interaction 
\begin{equation}
- g C^W_{\alpha \beta} \sqrt{2} v^2
(\bar{e^c}_\alpha \sigma^{\mu \nu} P_L \nu_\beta
+ \bar{\nu^c}_\alpha \sigma^{\mu \nu} P_L e_\beta )
[2 \partial_\mu W^+_\nu]  
\end{equation}
in the loop diagram on the LHS of figure (\ref{nonzero}),
gives
\beq
 C^M_{\alpha \beta}(m_W)  =  C^{M (0)}_{\alpha \beta} + 
%\left( 
  \frac{6 g^2 |m^{e2}_\alpha - m^{e2}_\beta| }{16 \pi^2}
%\frac{g}{\sqrt{2}} \right) \left( 
\frac{   [\mu]_{\alpha \beta}}{4 v^4 g  \sin \theta_W} 
%\right) 
\log 
\left( \frac{\Lambda_{NP}^2}{m_W^2} \right)
\label{presoln}
\eeq
where there is no sum on $\alpha, \beta$. 
Taking vacuum expectation values, this contributes
\beq
 \frac{1}{2} [ \delta m]_{\alpha \beta} = \langle C_M O_M 
\rangle_{\alpha \beta} 
 \simeq  
   C_{\alpha \beta}^{M (0)} v^4 + 
\left| \frac{m_\alpha^{e2} - m_{\beta}^{e2}}{m_\tau^2} 
\frac{[\mu]_{\alpha \beta}}{10^{-12} \mu_B} \right|
\log 
\left( \frac{\Lambda_{NP}^2}{m_W^2} \right)
 \times .014  ~ {\rm eV}
\label{soln}
\eeq
to the neutrino mass matrix.

Consider now the second scenario, where 
 the  New Physics only  produces  a
non-negligible Wilson coefficient for $O_B$.
Matching $O_B$ onto
the neutrino magnetic moment gives
\begin{equation}
 C^B_{\alpha \beta}(\Lambda_{\rm NP}) =  \frac{ [\mu]_{\alpha
    \beta}}{2 g' \cos \theta_W}
\end{equation}
 In general
$O_B$ can  mix into all possible dimension 7 $\Delta L = 2$ operators.
 For the purpose of studying the mixing into the neutrino mass
operator $O_M$,  we should calculate those divergent Green's functions
of $O_B$ which will mix into $O_M$.
We find that $O_B$ mixes into $O_W$,
 but not into
some other possibilities. 
We estimate the $O_B$ 
contribution to $O_M$ from its second order mixing through $
\hat{\gamma}^{(0)}_{WM}\hat{\gamma}^{(0)}_{BW}$:
\begin{equation}
  C_{\alpha \beta}^M(m_W) \sim  C_{\alpha \beta}^{M (0)} + 
 \frac{ 3 \tan^2 \theta_W  [\mu]_{\alpha \beta} |m^{e2}_\alpha -
  m^{e2}_\beta|}{4 v^4 e} 
  \left(\frac{g^2}{16 \pi^2}
  \log \left( \frac{\Lambda_{NP}^2}{m_W^2} \right) \right)^2
\label{est}
\end{equation}
If the magnetic moment is generated by the $O_B$ operator,
then its contribution to neutrino masses is  suppressed by
$\sim (\frac{\alpha}{4 \pi} \log )^2$. This 
 will not give interesting constraints on $[C^B]$.

%%%%%%%%%%%%%%%%%%%%%%%%%%%%%%%%%%%%%%%%%%%%%%%%%%%%%%%%%%%%%%%%%%%%%%
%% SECT DISC  %%%%%%%%%%%%%%%%%%%%%%%%%%%%%%%%%%%%%%%%%%%%%%%%%%%%%%
%%%%%%%%%%%%%%%%%%%%%%%%%%%%%%%%%%%%%%%%%%%%%%%%%%%%%%%%%%%%%%%%%%%%%%

\section{Discussion}
\label{sowhat}

In this section, we consider the phenomenological implications
of the magnetic moment contribution to the neutrino mass matrix.
We first review what is known about $\mnu$, then discuss
eqns (\ref{soln}) and (\ref{est}).

\subsection{observed parameters of the light $\nu$ sector}
The $\nu$ mass matrix in flavour space is
\beq
\mnu = U_{\alpha k} U_{\beta k} m_k
\eeq
where $\{ m_k \}$ are the neutrino masses, and 
$U$ is the MNS matrix, parametrized as
\bea U= \hat{U}\cdot {\rm diag}(1 ,e^{i\alpha} ,e^{i\beta}) ~~~.
\label{UV}
\eea
 $\alpha$ and $\beta$ are ``Majorana'' phases,
 and $\hat{U}$ has the form of the CKM matrix
\beq
 \label{Vdef} 
\hat{U}= \left[ \begin{array}{ccc}
c_{13}c_{12} & c_{13}s_{12} & s_{13}e^{-i\delta} \\
-c_{23}s_{12}-s_{23}s_{13}c_{12}e^{i\delta} & 
 c_{23}c_{12}-s_{23}s_{13}s_{12}e^{i\delta} & s_{23}c_{13} \\
s_{23}s_{12}-c_{23}s_{13}c_{12}e^{i\delta} & 
 -s_{23}c_{12}-c_{23}s_{13}s_{12}e^{i\delta} &
  c_{23}c_{13}  
\end{array} \right]  ~~~.
\eeq
Current data  \cite{moriond}
gives   $\theta_{23} \simeq \pi/4$, $\sin^2 \theta_{12} 
\simeq .29$,  $ \sin^2 \theta_{13} \lsim 0.035$, 
an atmospheric mass difference  $\Delta_@ m^2 \simeq (0.049 eV)^2$,
and a solar difference   $\Delta_\odot m^2 \simeq (0.0089 eV)^2$.
The absolute value of the mass scale in undetermined,
as is the ordering of the eigenvalues. It is convenient to
label the mass pattern  as degenerate
($m_i \gg \sqrt{\Delta m^2_{atm}}$), hierarchical
($m_3^2 \simeq \Delta m^2_{atm}, > m_2^2, m_1^2$)
or inverted 
($m_3^2 \ll m^2_{1,2}, ~~ m^2_{1,2} \simeq \Delta m^2_{atm}$).

For comparison with the magnetic moment contributions
to the mass matrix, it will be useful to have an
idea of the numerical values of $\mnu$.
Taking  $\theta_{12} \simeq 0.55$, $\theta_{23} \simeq \pi/4$
$m_1 = 0$, $|m_3| = .049$ eV and $k_2 = e^{2 i (\alpha - \beta)}m_2/m_3$
$(|k_2| =  0.18$),
in agreement with the hierarchical interpretation of current
data, gives a mass matrix  

\beq
 \label{mnum2} 
[m_\nu]\simeq \left[ \begin{array}{ccc}
.30 k_2  & 
.32 k_2  + .35s &
 -.32 k_2  + .35s \\
.32 k_2  + .35s &
.35 k_2  + .25 &
-.35 k_2  + .25 \\
 -.32 k_2  + .35s &
-.35 k_2  + .25  &
.35 k_2  + .25
\end{array} \right]~~ \times .1 {\rm eV}  ~~~.
\eeq
where the phase $\delta$ has been
absorbed into the unknown angle $ s = \sin \theta_{13} e^{-i \delta}$.
From data,  $|s| \leq 0.2$, so  we took $\cos \theta_{13} = 1$, and
dropped terms of order $s k_2$.

\subsection{the magnetic moment contribution}

The magnetic moment contribution to the neutrino
mass matrix, from eqn (\ref{soln}), reads 
\beq 
[ \delta m_\nu]\simeq   b \left[ \begin{array}{ccc}
0 & 0.004 \tilde{\mu}_{e \mu} &    \tilde{\mu}_{e \tau} 
 \\
0.004  \tilde{\mu}_{e \mu}  &
0 &   \tilde{\mu}_{\mu \tau} 
 \\
  \tilde{\mu}_{e \tau}  &
   \tilde{\mu}_{\mu \tau}  &
0
\end{array} \right]~~  %\times \left(  \frac{a}{10^{-12} \mu_B} \right)
\times .1 {\rm eV}  ~~~, 
\label{m}
\eeq
where the $\tilde{\mu}_{\alpha \beta}$ are
the magnetic moments measured in units of
$10^{-12} \mu_B$, and the
log was conservatively estimated by
taking  $\Lambda_{NP} \sim$ 1 TeV.
If the magnetic moment is generated by $O_W$,
then $b = 1$. However, if $O_B$  is the magnetic
moment operator, then
%NEW
 $b %\sim 2 \alpha'/(4 \pi) \log (\Lambda_{NP}/m_W )
\sim \alpha_{em}/\pi$,
 and the contribution to the mass matrix is reduced.

The mass matrix (\ref{m}) by itself  is not phenomenologically viable,
because it imposes $\sum_i m_i = 0$, and predicts
relations between the mixing angles and mass differences
(it has only three free parameters). However, it naturally
gives large mixing angles, so could contribute to $[m_\nu]$
in conjunction with some other (flavour diagonal?) source
of neutrino masses \cite{aj,cei}.
For the case of  degenerate neutrinos, there are no bounds on $[\mu]$,
in general. However,  if the main 
source of neutrino masses is
flavour diagonal, then the limits discussed below   also apply.

The phenomenological consequences of
eqn (\ref{m}) can be seen by considering a series
of cases:
\begin{enumerate} 
\item  suppose that the magnetic moment operator
is $O_W$, so
$[\mu] = 4 v^2 e [C^W]$ and 
 $b = 1$ in eqn (\ref{m}).
\begin{itemize}
\item If  $[\mu]_{\mu \tau} \gappeq
10^{-12} \mu_B$, then for
hierarchical or inverted neutrino masses (but not
for degenerate),   we find
\footnote{For simplicity, we present bounds on the matrix elements
of $[\mu]$, however in reality they  apply to $[C^W]$.}
  $[\delta m_\nu]_{\mu \tau} \gappeq
[m_\nu]_{\mu \tau} $.  So either  $[\mu]_{\mu \tau} \leq
3 \times 10^{-13} \mu_B$, or there is some mild cancellation
between $[\delta m_\nu]$ and other sources to $[m_\nu]$.
The numerical ``bound'' is fuzzy because it
depends logarithmically on $\Lambda_{NP}$, and
slightly on the mass pattern and phases of $[m_\nu]$.
\item  now suppose that  $[\mu]_{e \tau} \sim
10^{-12} \mu_B$. If the neutrino masses are inverted,
it is similar to the previous case. However, for
hierarchical neutrino masses,  $$
\frac{[\delta m_\nu]_{e \tau}} 
{[m_\nu]_{e \tau}} \sim 10$$
 So    $[\mu]_{e \tau} <
10^{-13} \mu_B$, which  is too small to have an effect in
solar physics, or there is a significant cancellation
between $[\delta m_\nu]_{e \tau}$ and $ [m_\nu]_{e \tau}$.
 It is usual, in quoting bounds on the coefficients
of non-renormalizable operators, to suppose that
only one operator at a time is present and neglect the possibility
of cancellations. 
In this approach, the bound on the  coefficient $[C_W]_{e \tau}$
corresponds to     $[\mu]_{e \tau} <
10^{-13} \mu_B$.

\item the contribution of  $[\mu]_{\mu e}$ to
the neutrino mass matrix is always small (suppressed
by $m_\mu^2$),  so if there is time-dependence in
the solar neutrino signal, it is more likely due to 
  $[\mu]_{\mu e}$ .
\end{itemize}
\item Now consider the case where the neutrino
magnetic moment is due to the operator $O_B$. The
contribution to $[m_\nu]$  is of order
$(\frac{\alpha}{4 \pi} \log)^2$, so in eqn (\ref{m}), $b \sim \alpha_{em}/\pi $, 
and  $[ \delta m_\nu]_{\alpha \beta} < [m_\nu]_{\alpha \beta}$
for magnetic moments of order $[\mu]_{\alpha \beta} \sim
10^{-12} \mu_B$. So the coefficients $[C^B]$
are more tightly constrained by the upper bounds
on magnetic moments, than they are by their 
loop contributions to $[m_\nu]$. 

Notice,however, that %for hierarchical neutrino masses,
\footnote{This relies on the rough estimate of 
$b$ from eqn (\ref{est}).}
$ [{\delta} m_\nu]_{e \tau}  \simeq 
3 \times 10^{-4} \left( \frac{[\mu]_{e \tau}}{10^{-12} \mu_B} \right)$  eV, 
and for  a hierarchical neutrino mass pattern:
$m_{e \tau} \simeq 5 \times 10^{-3}$ eV.  So 
a magnetic moment of order  $[\mu]_{e \tau}  \sim 10^{-10} \mu_B$
(less than the experimental bound, but
exceeding the astrophysical one)
contributes a larger loop correction to $[m_\nu]_{e \tau}$
than its measured value.
\end{enumerate}

%\subsection{back to motivation}

The original hope  was to 
``reconstruct'' New Physics in the lepton sector. 
This is straightforward if the new particles
and the form of their  interactions are known; 
one can then attempt to extract the numerical
value of the new coupling constants from the coefficients
of operators involving SM fields 
\footnote{For instance, in
the SUSY  type-1 seesaw, one can in principle extract the
heavy RH neutrino masses and the neutrino
Yukawa matrix, from the mass matrices of the 
sneutrinos and the light neutrinos \cite{di1}.}.
However,  the question  is  whether 
one can also learn about the New Physics ``mechanism'',
that is, the particle content and type of
interactions.  We have not attempted to
do this, but the result of this paper 
emphasizes the confusion: if the new physics 
scale is low enough to generate observable magnetic
moments, then neutrino masses could arise from
the usual dimension five operator, or the dimension
seven operator $O_M$, and these are indistinguishable.

\subsection{summary}
Transition (flavour changing)
 magnetic moments among Standard Model neutrinos 
are lepton number violating, so  they  contribute,
via SM loop effects, to majorana masses. This is
similar to the remark that New Physics inducing
neutrinoless double beta decay necessarily contributes
to majorana masses \cite{vs}. The transition magnetic
moment matrix  $[\mu]$ is also flavour {\it antisymmetric}, so
must be multiplied by charged lepton masses in the SM loops
contributing to the flavour {\it symmetric} majorana
mass matrix $[m_\nu]$. The largest contribution of
  $[\mu]$  to   $[m_\nu]$  is therefore in the third
generation.

 Transition magnetic moment operators  in the SM
have mass dimension $\geq 7$,
where there are two possible operators $O_B$ and $O_W$
(eqn \ref{magmoSM}).
$O_W$ contributes at one-loop to  $[m_\nu]$, and
$O_B$ at two-loop (see {\it e.g.} fig. \ref{nonzero}).
We can set upper bounds on the coefficient matrices $[C^W]$ and
$[C^B]$
of these operators, from requiring that their
loop contributions to $[m_\nu]$ be small enough;
these bounds then constrain the magnitude of the
magnetic moment that the operator can generate.
The resulting bounds on  $[C^B]$ are weaker than current
experimental limits on neutrino transition magnetic
moments ($\sim 10^{-10} \rightarrow
3 \times 10^{-12} \mu_B$, see eqns \ref{bdexpt}  and \ref{bdastro}); 
that is, the non-observation of
neutrino transition magnetic moments sets tighter bounds
on the coefficients of this operator than their loop
contributions to neutrino masses.
However, the case of $O_W$ is somewhat different.  

If the light neutrino masses are non-degenerate, then
the bound on $[C^W]_{\alpha \tau}$ from its contribution
to $[m_\nu]$ , translates into the limit
$[\mu]_{\alpha \tau} \lappeq 3 \times 10^{-13} \mu_B$. 
If the light neutrino masses are hierarchical, then
we obtain 
$[\mu]_{e \tau} \lappeq  10^{-13} \mu_B$. Notice, however,  that these bounds
apply to the coefficients of the operator $O_W$,
although for simplicity we quote them as limits on
the neutrino transition magnetic moment. 
A larger $[\mu]_{\alpha \tau}$ could be consistent
with the observed masses, if, for instance, it
was generated by $O_B$.

 \subsection*{Acknowledgements}
At the early stages of this project, 
the work  of S.D. was supported by  a PPARC
Advanced Fellowship. The work of A.S. has been supported by the 
Spanish MCyT under the Grant
BFM2002-00568.

\subsection*{note added}
In a recent paper \cite{Bell},  Bell  et al perform a
similar calculation for Dirac neutrinos.
They calculate bounds  
$$
\mu_B \lappeq 10^{-14} \mu_B
$$
on Dirac  magnetic moments,  from their  
contributions to Dirac neutrino masses. 
 
The Dirac  magnetic moment is not required to
be  flavour antisymmetric, so  its contribution
to  neutrino masses is not suppressed by 
Yukawa couplings. This allows a stronger bound,
which applies to all flavours.
This suggests that only the majorana magnetic moment
$\mu_{e \mu}$ can be large enough to affect
solar physics.

\end{document}